\documentstyle[12pt]{article}
\newcommand{\heading}[1]{\vspace*{15mm}
{\Large\begin{center} {\bf{#1}} \end{center}}}
\renewcommand{\author}[3]{\vspace{5mm}
\begin{center}
{\normalsize \rm #1}\\    
{\normalsize \it #2}\\    
{\normalsize \it #3}\\    
\vspace{0.65cm}\framebox[3.4truecm]{\rule[-1.9cm]{0.cm}{3.4truecm}}
\vspace{0.35cm}\end{center} }

\def\lsim{~\rlap{$<$}{\lower 1.0ex\hbox{$\sim$}}}
\def\gsim{~\rlap{$>$}{\lower 1.0ex\hbox{$\sim$}}}

\newcommand{\mn}{{\em Mon. Not. R. astr. Soc}}
\newcommand{\apj}{{\em Astrophys. J.}}
\newcommand{\aj}{{\em Astron. J.}}
\renewcommand{\aa}{{\em Astr. Astrophys.}}

\newcommand{\nat}{{\em Nature}}
\newcommand{\et}{{\em et al.}\,\,}

\newcommand{\rmsub}[2]{#1_{\rm #2}}
\newcommand{\dns}{\mbox{$\rmsub{D}{n} - \sigma \:$}}

\begin{document}
\heading{WHAT DO WE MEAN BY `MALMQUIST BIAS'?}

\author{M. A. HENDRY$^{1}$, J. F. L. SIMMONS$^{2}$, A. M. NEWSAM$^{2}$}
       {$^{1}$ Astronomy Centre, University of Sussex, Brighton, BN1 9QH, UK.}
       {$^{2}$ Dept. of Physics and Astronomy, University of Glasgow, Glasgow
       G12 8QQ, UK.}

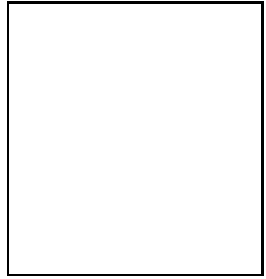
\begin{abstract}{\baselineskip 0.4cm
In this paper we review the two main approaches to the problem of
Malmquist bias which have been adopted in the cosmology literature, and show
how these two formulations of the problem represent fundamentally different
views of the nature of probability. We discuss the assumptions upon
which both approaches are based and indicate some of their limitations.
In particular we identify a basic flaw in the definition of
homogeneous and inhomogeneous Malmquist corrections as they have frequently
been applied in the literature, and indicate how this flaw may be corrected.}
\end{abstract}

\clearpage
\newpage
\section{Introduction}
\label{sec:intro}
\setlength{\baselineskip}{15pt}
In recent years the analysis of the peculiar velocity field using
redshift-independent galaxy distance indicators has significantly enhanced
our understanding of the formation and evolution of large scale structure.
The most prevalent examples of these indicators have been the Tully-Fisher
(TF) and \dns relations, and their use in e.g. the {\sc potent} analyses has
contributed to the solid body of evidence in support of coherent streaming
motions over scales on the order of 100Mpc \cite{pot90}, \cite{potiras} --
evidence which, nevertheless, has attracted considerable controversy in the
literature, not least because of the difficulties which it presents for
popular theories of structure formation. Much of this debate has focussed upon
the statistical properties of the TF and \dns relations and the
issue of how best to deal with the systematic errors which arise when they
are applied to surveys which are subject to observational selection. These
systematic effects have been referred to generically in the literature as
`Malmquist bias'. There exists, however, a great deal of confusion over
precisely what is meant by Malmquist bias (or the `M word' as it has been
labelled at this conference!). Different authors have
used the term `Malmquist bias' to denote different -- and often contradictory
-- effects \cite{ls}, \cite{lb}, \cite{bick}. It is not surprising, therefore,
that no consensus has yet been reached about how best to eliminate Malmquist
bias in studying the peculiar velocity field.

In this paper we examine the statistical basis of the different approaches to
the problem of Malmquist bias which
have been adopted in the literature. We discuss the model assumptions upon
which each depends, and the extent to which these assumptions may be
generalised. We thus indicate how one may formulate a rigorous, consistent
treatment of the problem of galaxy distance estimation and Malmquist bias.

\section{Unbiased distance estimators}
\label{sec:unbias}
\baselineskip 0.6cm
We have already heard from Brent Tully at this meeting a number of
adjectives which he used to describe previous approaches to the problem of
Malmquist bias in the literature. We can now add two further terms to this
list: {\em frequentist \/} and {\em Bayesian \/}. Examples of where the
former approach has been adopted in recent literature include \cite{hs},
\cite{hs93}, \cite{bick}, \cite{pt}, \cite{st}: this approach follows closely
the original treatment of the statistical effect by Malmquist \cite{malm}.
The approach which we may categorise as Bayesian, on the
other hand, has been adopted in e.g. \cite{ls}, \cite{lb},
\cite{pot90}, and is, in fact, closer in spirit to early work by Eddington,
\cite{edd}, on correcting observational errors. (Indeed Lynden-Bell also refers
to Eddington-Malmquist bias, recognising the true origins of his
statistical approach to the effect). Our adoption of the terms frequentist
and Bayesian reflects the fact that at the heart of the difference between
these two approaches lies the long-standing dichotomy between a Bayesian and
a frequentist view of the nature of probability, as we will now explain.

\subsection{A `frequentist' approach to bias}
\label{sec:freq}

The `frequentist' view of the nature of probability is essentially based on
the intuitively familiar idea that the probability of an
event is a measure of the {\em relative frequency \/} of that event occurring
in a large number of repeated experiments. In the limit as the number of
experiments tends to infinity the frequency histogram reduces to the
probability
density function (pdf) of a random variable. In the present context our
`experiment' is the estimation of the distance to a given galaxy -- e.g. by
measuring its apparent magnitude and 21cm line width in the case of the
TF relation. Crucial to the frequentist approach is the idea that
the true distance of this galaxy is a fixed, though of course unknown,
parameter -- an `unknown state of nature' in statistical parlance.

We can state these ideas more rigorously as follows. Suppose we are estimating
the distance to a given galaxy which lies at true distance, $\rmsub{r}{0}$.
Let ${\bf \hat{r}}$ denote a galaxy distance estimator. For the TF
relation ${\bf \hat{r}}$ is a function of apparent
magnitude and 21cm line width -- i.e. ${\bf \hat{r}} = f({\bf m},{\bf P})$.
(Here we have introduced several items of notation. We follow the standard
statistical practice of denoting a random variable by a bold face character,
and an estimator of a parameter by a caret. We also adopt ${\bf P}$ as a
shorthand for log(line width) -- a notation used by several previous authors
\cite{bick}, \cite{teer}).
The precise form of this function depends upon the joint distribution of
${\bf m}$ and ${\bf P}$, and also how the TF relation is
calibrated, as we discuss below.

Consider now the pdf, $p({\bf \hat{r}} | \rmsub{r}{0})$, of ${\bf \hat{r}}$,
conditional upon the true distance of the galaxy -- the parameter which we
are estimating. ${\bf \hat{r}}$ is defined to be {\em unbiased \/}
(c.f. \cite{ks}, \cite{mg}) if the mean, or {\em expected \/}, value of
${\bf \hat{r}}$ is equal to the true distance, $\rmsub{r}{0}$, of the galaxy.
More generally the bias, $B({\bf \hat{r}},\rmsub{r}{0})$, of ${\bf \hat{r}}$ is
defined as:-
\begin{equation}
\label{eq:biasdefn}
   B({\bf \hat{r}},\rmsub{r}{0}) = E({\bf \hat{r}} | \rmsub{r}{0})
   - \rmsub{r}{0} =
   \int {\bf \hat{r}} p({\bf \hat{r}} | \rmsub{r}{0})
   d {\bf \hat{r}} - \rmsub{r}{0}
\end{equation}
Equation \ref{eq:biasdefn} immediately demonstrates the importance of finding
an unbiased distance estimator in this approach, since we see that the bias
of ${\bf \hat{r}}$ is a function of the (unknown) true distance,
$\rmsub{r}{0}$. We should also note that we can define in an analogous
fashion the bias of an estimator of {\em any \/} parameter: in particular an
estimator of the true log distance of a galaxy.

Having thus defined what we mean by the bias of an estimator we now consider
how this bias arises in estimating galaxy distances, taking again as our
example the TF relation.

\subsection{Malmquist bias in the Tully-Fisher relation}
\label{sec:malm}

Let ${\bf M}$ denote the absolute magnitude of the given galaxy. Ignoring
absorption and cosmological effects, the following equation holds:-
\begin{equation}
\label{eq:mMrel}
   {\bf m} = {\bf M} + 5 \log \rmsub{r}{0} + 25
\end{equation}
where the true distance, $\rmsub{r}{0}$, of the galaxy is in Mpc. From equation
\ref{eq:mMrel} an obvious estimator of log distance is given by:-
\begin{equation}
\label{eq:logrhat}
   {\bf \widehat{\log r}} = 0.2 ( {\bf m} - {\bf \hat{M}} - 25 )
\end{equation}
where ${\bf \hat{M}}$ is some estimator of the galaxy's absolute magnitude.
In early studies of the peculiar velocity field \cite{rub}, \cite{st75} a
fixed fiducial value for ${\bf \hat{M}}$ was adopted from prior
considerations -- i.e. assuming that the observed galaxies were standard
candles. The TF relation provides a better estimator of ${\bf M}$ by making
use of the strong observed correlation between the luminosity and 21cm line
width of spirals. The relation is usually calibrated by performing a linear
regression on a calibrating sample of galaxies whose distances are otherwise
known. Thus we obtain a linear relationship between ${\bf \hat{M}}$ and
${\bf P}$, viz:-
\begin{equation}
\label{eq:Mhat}
   {\bf \hat{M}} = \alpha {\bf P} + \beta
\end{equation}
where $\alpha$ and $\beta$ are constants.

The choice of which linear regression is most appropriate is
non-trivial, however, particularly when one's survey is subject to
observational selection effects. We can illustrate this with the following
simple example. Suppose that the intrinsic joint distribution of absolute
magnitude and log(line width) is a bivariate normal. The left hand panel of
Figure \ref{fig:TF1} shows schematically the scatter in the TF relation in
this case, for a calibrating sample which is free from selection effects --
e.g. a nearby cluster. (More precisely, the ellipse shown is an
isoprobability contour enclosing a given confidence region for ${\bf M}$ and
${\bf P}$). The solid and dotted lines show the linear relationship obtained by
regressing line widths on magnitudes and magnitudes on line widths
respectively.
Thus the dotted line is defined as the expected value of ${\bf M}$ at given
${\bf P}$, while the solid line is defined as the expected value of ${\bf P}$
at given ${\bf M}$. Since in practice one wishes to infer the value of
${\bf M}$ from the measured value of ${\bf P}$, the ${\bf M}$ on ${\bf P}$
regression is often referred to in the literature as defining the `direct' TF
relation, while using the ${\bf P}$ on ${\bf M}$ regression defines the
`inverse' TF relation. For the bivariate normal case the equations of the
direct and inverse regression lines are as follows:-
\begin{equation}
\label{eq:MgivP}
   E ({\bf M}|{\bf P}) = \rmsub{M}{0} + \frac{\rho \rmsub{\sigma}{M}}
                        {\rmsub{\sigma}{P}} ({\bf P} - \rmsub{P}{0})
\end{equation}
\begin{equation}
\label{eq:PgivM}
   E ({\bf P}|{\bf M}) = \rmsub{P}{0} + \frac{\rho \rmsub{\sigma}{P}}
                        {\rmsub{\sigma}{M}} ({\bf M} - \rmsub{M}{0})
\end{equation}
where $\rmsub{M}{0}$, $\rmsub{P}{0}$, $\rmsub{\sigma}{M}$, $\rmsub{\sigma}{P}$
and $\rho$ denote the means, dispersions and correlation coefficient of the
bivariate normal distribution of ${\bf M}$ and ${\bf P}$.
Both regression lines can be written in the form of equation \ref{eq:Mhat},
thus defining ${\bf \hat{M}}$ as a function of ${\bf P}$, although of
course the constants $\alpha$ and $\beta$ will be different in each case.
Moreover the definition of ${\bf \hat{M}}$ is subtly different in each case.
For the direct regression ${\bf \hat{M}}$ is identified as the mean
absolute magnitude at the observed log line width. For the inverse regression
on the other hand ${\bf \hat{M}}$ is defined such that the observed log line
width is equal to its expected value when ${\bf M} = {\bf \hat{M}}$.
Consequently, as is apparent from their slopes, the direct and inverse
regression lines give rise to markedly different distance
estimators, although it is straightforward to show that in the absence of
selection effects both estimators are unbiased.

The situation is very different when we include the effects of observational
selection, however. This is illustrated in the right hand panel of
Figure \ref{fig:TF1}, which
shows the scatter in the TF relation in a calibrating sample subject to a
sharp cut-off in absolute magnitude -- as would be the case in e.g. a
distant cluster observed in an apparent magnitude-limited survey. We can
see that in this case the slope of the direct regression
of ${\bf M}$ on ${\bf P}$ is substantially changed from that in the nearby
cluster -- indeed the direct regression is no longer linear at all. This means
that if one calibrates the TF relation in the nearby cluster using the direct
regression and then applies this relation to the more distant cluster, one
will systematically underestimate its distance, since the expected value of
${\bf M}$ given ${\bf P}$ in the distant cluster is systematically brighter
than that in the nearby cluster as fainter galaxies progressively `fade out'
due to the magnitude limit. The corresponding `direct', or `M on P', log
distance estimator -- obtained by substituting the appropriate constants into
equation \ref{eq:Mhat} and then equation \ref{eq:logrhat} -- will therefore
be negatively biased. This is precisely analogous to the bias identified by
Malmquist \cite{malm} in considering the mean absolute magnitude of standard
candles in a sample with a sharp apparent magnitude limit.
\begin{figure}
\noindent
\vspace{9cm}
\caption{Schematic Tully-Fisher relations for the case of a nearby, completely
sampled, cluster and a distant cluster subject to a sharp selection limit on
absolute magnitude}
\label{fig:TF1}
\end{figure}

\subsection{Properties of `Schechter' estimators}
\label{sec:schech}

In an important paper \cite{sch} Schechter observed that the slope of
the inverse regression line is unchanged, irrespective of the completeness
of one's sample, provided that the selection effects are in magnitude only.
We can see that this observation is valid in the simple case considered in
Figure \ref{fig:TF1}. In other words the expected value of ${\bf P}$ given
${\bf M}$ is unaffected by the Malmquist effect and, therefore, defines an
unbiased log distance estimator.
Although the unbiased property of the inverse regression line has been
generally recognised (c.f. \cite{pt}, \cite{tull}, \cite{bick}, \cite{lb}), its
ramifications for estimating galaxy distances have not been fully appreciated.

We have carried out an extension of Schechter's ideas to more realistic
situations \cite{h92}, \cite{hs93} and examined the extent to which the
assumptions upon which they are based may be generalised. We now briefly
summarise the properties of unbiased `Schechter' estimators.

\begin{enumerate}
\item{In a sample subject to observational selection effects, it is possible
to define a linear estimator of log distance
which is unbiased at all true distances -- provided that the following two
conditions are met: the measurements of line width are free from selection
effects and the conditional expectation of log(line width) at given absolute
magnitude is linear in {\bf M}. Moreover, the appropriate linear combination
of ${\bf M}$ and ${\bf P}$ corresponds exactly to the estimator derived from
the inverse regression of line widths on magnitudes, as prescribed in
\cite{sch}.}

\item{The `Schechter' estimator is the {\em only \/} unbiased linear
estimator. Any other linear combination of magnitude and line width, and in
particular any other regression line, yields an estimator which is biased at
all true distances for a magnitude selection function. Examples of biased
regression lines in this case include not only the direct or 'M on P'
regression shown above (c.f. \cite{lb}, \cite{bick}) but also the orthogonal
\cite{gir}, bisector \cite{pt} and mean \cite{mould} regression lines.}

\item{The shape of the pdf, and hence in particular the {\em variance \/} of
the Schechter estimator, is constant at all true distances and is in fact
identical to that of the intrinsic conditional distribution of ${\bf P}$
given ${\bf M}$. This conditional distribution is frequently modelled to be
gaussian, thus implying that the Schechter estimator is gaussian in this
case. Again the Schechter estimator is unique in this regard - for any other
general linear estimator the shape of
its distribution becomes distorted at large true distances, as the effects of
selection become significant. It follows from this property that confidence
intervals derived from the Schechter estimator will have constant width
\cite{hs}.}

\item{The unbiasedness of the Schechter estimator holds for an arbitrary
luminosity function. This is a particularly useful property, since the bias
of any other linear estimator will depend explicitly upon the form of the
luminosity function, so that any attempt to remove the bias would necessarily
be model dependent \cite{bick}.}

\item{One may also define an unbiased log distance estimator for other distance
indicators, including the \dns and magnitude-colour relations, subject to the
same condition that there be one observable free from selection, but
{\em not \/} requiring one observable to be distance-independent. In a
diameter-complete survey, for example, one may construct an unbiased distance
estimator from the observed angular diameter and apparent magnitude. As above,
it is easy to show that this unbiased estimator corresponds exactly to
the regression of the selection-free observable upon the other observable.}

\item{If there is {\em no \/} selection-free observable, then an unbiased
estimator {\em cannot \/} be defined as a simple linear combination of the
observables. In the context of both the Tully-Fisher and \dns relations,
however, this is somewhat less of a problem than one might expect. Most
surveys will be subject to a {\em lower \/} selection limit on line width or
velocity dispersion. This selection limit becomes increasingly {\em less \/}
important at larger true distances, however. This is easy to understand,
since at large distances only intrinsically brighter (or larger), and thus
sufficiently large line width, galaxies will be observable \cite{ls}.
It is found that at cosmologically interesting distances, the Schechter
estimator is still effectively unbiased in this case.}
\end{enumerate}

\subsection{A Bayesian approach to unbiased distance estimators}
\label{sec:bayes}

We now consider the second, essentially Bayesian, approach to the
problem of Malmquist bias, adopted in e.g. \cite{lb}, \cite{lb}, and
\cite{potiras}. The crucial difference in this approach is that the true
distance, ${\bf \rmsub{r}{0}}$ of a galaxy is itself regarded as a random
variable. Hence one must assign a prior probability distribution for
${\bf \rmsub{r}{0}}$, based upon an assumed spatial density distribution
and selection function. Following the measurement of some distance
estimator, ${\bf \hat{r}}$ for each galaxy -- using e.g. the TF or \dns
relation -- one can define a posterior distribution for
${\bf \rmsub{r}{0}}$ which will differ from the prior. It is the properties
of this posterior distribution which are considered in defining an
estimator as unbiased. We can use Bayes' theorem to derive an expression for
the posterior distribution, $p({\bf \rmsub{r}{0}} | {\bf \hat{r}})$, viz:-
\begin{equation}
\label{eq:bayes}
   p ({\bf \rmsub{r}{0}} | {\bf \hat{r}}) =
   \frac{ p ({\bf \hat{r}} | {\bf \rmsub{r}{0}}) p({\bf \rmsub{r}{0}})}
     {\int p({\bf \hat{r}} | {\bf \rmsub{r}{0}}) p({\bf \rmsub{r}{0}})
           d {\bf \rmsub{r}{0}}}
\end{equation}
Here $p({\bf \rmsub{r}{0}})$ is the prior distribution for ${\bf \rmsub{r}{0}}$
and $p ({\bf \hat{r}} | {\bf \rmsub{r}{0}})$ is known as the likelihood
function, and is in fact simply the pdf of our distance estimator,
${\bf \hat{r}}$, as discussed previously.

In this approach the distance estimator, ${\bf \hat{r}}$, is defined as
unbiased if the expected value of ${\bf \rmsub{r}{0}}$
with respect to this posterior distribution,
$p({\bf \rmsub{r}{0}} | {\bf \hat{r}})$, is equal to ${\bf \hat{r}}$.
In general the bias of ${\bf \hat{r}}$ is defined as:-
\begin{equation}
\label{eq:baybias}
   B({\bf \hat{r}},{\bf \rmsub{r}{0}}) = E({\bf \rmsub{r}{0}} | {\bf \hat{r}})
   - {\bf \hat{r}} =
   \int {\bf \rmsub{r}{0}} p ({\bf \hat{r}} | {\bf \rmsub{r}{0}})
   d {\bf \rmsub{r}{0}} - {\bf \hat{r}}
\end{equation}
Compare this expression with equation \ref{eq:biasdefn} above.
By assuming a prior distribution and likelihood function we can see from
equations \ref{eq:bayes} and \ref{eq:baybias} that one may derive a
Malmquist correction to remove the bias of a `raw' distance estimator, so
that the corrected estimator is unbiased.

Of course one may formulate this approach for the corresponding unbiased
estimator of log distance in an analogous fashion. Lynden-Bell \et \cite{lb}
do precisely this, assuming a prior distribution which corresponds to a
constant spatial number density of galaxies and assuming for
their raw log distance estimator a gaussian pdf of mean value equal to the true
log distance and of constant variance. These assumptions imply a constant,
or {\em homogeneous \/} Malmquist correction: in other words all raw distance
estimates are rescaled by a constant factor. Clearly this assumed prior
will be incorrect -- due to both galaxy clustering and observational selection
effects -- although it might be regarded as a reasonable first approximation.

Landy and Szalay \cite{ls} present an improved treatment by explicitly
recognising the Bayesian nature of this problem and proposing that
one use the observed distribution of raw distance estimates to provide a
better approximation to the prior distribution of log true distance. In
principle this prior would take account of the effects of clustering and
selection which render the observed distribution inhomogeneous -- thus
leading to the definition of an {\em inhomogeneous \/} Malmquist correction.
They still assume, however, that the pdf of their raw log distance estimator
is gaussian, with constant variance and mean equal to the true log distance.

\section{Comparing the two approaches}
\label{sec:comp}

One might regard the differences between the two approaches we have outlined
for
interpreting, and dealing with, the effects of Malmquist bias as of no more
than semantic importance. This is far from the case, however. Firstly it is
worth pointing out that -- however valid the two approaches may be when
considered individually -- viewed together they are mutually inconsistent.
In other words an estimator defined as unbiased in the frequentist sense
{\em must always \/} be biased in the Bayesian sense, and vice versa. Hence any
analysis which is not self-consistent in its approach to bias throughout
will in general wind up deriving which are unbiased in neither the frequentist
nor the Bayesian sense!

Another important difference between the two approaches stems from the
dependence of the Bayesian description upon the assumption of a prior true
distance distribution. This results in a different Malmquist correction
for field and cluster samples \cite{lb}. This dichotomy can lead to ambiguity
in the case where cluster membership is unresolved -- as is frequently the case
with spiral galaxies. No such difficulty exists with the frequentist
description, however. Since the bias of a distance estimator is defined
conditionally upon the true (log) distance, it is easy to show \cite{h92}
that the definition of an unbiased distance estimator in this approach is
completely independent of the local galaxy number density. It would seem that
this important distinction has not been well appreciated in the literature.

In section \ref{sec:bayes} we noted that the homogenous and inhomogeneous
Malmquist corrections were both derived on the assumption that our raw log
distance estimator has a gaussian pdf with mean value equal to the true
log distance. In other words this means that the raw log distance estimator
is assumed to be unbiased -- according to the frequentist definition of
equation \ref{eq:biasdefn}. As we saw in section \ref{sec:schech},
{\em only \/} the Schechter log distance estimator, corresponding to the
inverse regression of line widths on magnitudes in the case of the
TF relation, has this property -- and even then only when one's sample is
free from line width selection effects. If any other regression line is
used to derive the raw distance estimator, then the assumptions
underlying the definition homogeneous and inhomogeneous Malmquist corrections
will no longer be valid, and the Malmquist corrections derived from the
raw distance estimator will {\em not \/} remove the effects of Malmquist bias
-- even in the (unlikely!) case where the prior true distance distribution is
known exactly.

In short, therefore, Malmquist corrections {\em must \/} be computed using a
Schechter log distance estimator as one's raw estimator if they are to be
in any way effective. This crucial result has not been recognised in the
literature, and to our knowledge every application of homogeneous and
inhomgeneous Malmquist corrections has been carried out using a biased raw
log distance estimator (c.f. \cite{pot90}, \cite{potiras}, \cite{lb},
\cite{ls}, \cite{df}).

We examine elsewhere in these proceedings \cite{nsh} the specific
implications of this important result for velocity field reconstruction
techniques -- in particular the {\sc potent} method \cite{pot90}.

\section{Summary}
\label{sec:summ}

In this paper we have seen how one may address the problem of Malmquist bias
in two distinctly different ways, essentially reflecting a frequentist
and Bayesian view of the nature of probability. We have shown how, following
either approach, one may in principle derive unbiased distance estimators
and have discussed the assumptions upon which this result holds. In the
frequentist approach the unbiased distance estimator corresponds to the
inverse regression of line widths on magnitudes -- as prescribed by Schechter
\cite{sch}. In the Bayesian picture unbiased estmators are defined by
computing the appropriate Malmquist correction to a raw distance estimator,
assuming a prior distribution for true distance. We have thus
identified a serious error in the application of homogeneous and inhomogeneous
Malmquist corrections in the literature, since these have been computed with
a biased `raw' distance estimator -- violating a basic assumption in their
definition. We have indicated how one may compute the proper Malmquist
corrections by using the Schechter estimator as one's raw distance estimator.

The question of which of these two approaches to the problem of Malmquist bias
is best has no clear-cut answer. We discuss some of the issues in more detail
elsewhere \cite{nsh}, \cite{hs93}. It suffices to say here that the main
requirement of any statistical analysis of galaxy distance estimates is to be
consistent. One should point out, however, that there are often
circumstances where one does not have complete freedom to choose either the
frequentist or Bayesian approach.

A common example of how this difficulty can arise is when one's velocity field
data must be heavily smoothed -- as is the case with the {\sc potent} method
\cite{pot90}. {\sc potent} requires the computation of a smoothed radial
peculiar velocity field at all points on a spherical grid, and accomplishes
this by using very large smoothing windows, of effective radius
$\sim$ 5000kms$^{-1}$. In interpolating a peculiar velocity from galaxies
appearing in the catalogue to a given spatial grid point, the
{\em essential \/} effect of the smoothing window is to pick out the galaxy
whose {\em estimated \/} position lies closest to the
prescribed grid point. The actual distance of this galaxy may be radically
different from its estimated distance, and will depend upon the true spatial
distribution of galaxies. In requiring that the mean smoothed radial peculiar
velocity be equal to the true radial velocity at that point, one finds that we
require equation \ref{eq:baybias} to vanish -- i.e. we want an distance
estimator which is unbiased according to the Bayesian description, thus
requiring the application of inhomogeneous Malmquist corrections. Of course,
as we have pointed out above, these corrections will be seriously inadequate
if one does not use a Schechter raw distanmce estimator -- a fact which does
not appear to have been realised by the {\sc potent} authors \cite{pot90},
\cite{potiras}.

We discuss elsewhere in some detail the effects of bias
and smoothing procedures on velocity field reconstructions with {\sc potent}
\cite{nsh}, \cite{snh}.

\vfill
\end{document}